\def\etal{{\it et al.}}

\def\spose#1{\hbox to 0pt{#1\hss}}
\def\lta{\mathrel{\spose{\lower 3pt\hbox{$\mathchar"218$}}
     \raise 2.0pt\hbox{$\mathchar"13C$}}}
\def\gta{\mathrel{\spose{\lower 3pt\hbox{$\mathchar"218$}}
     \raise 2.0pt\hbox{$\mathchar"13E$}}}
\def\ge{\mathrel{\spose{\lower 3pt\hbox{$-$}}
     \raise 2.0pt\hbox{$\mathchar"13E$}}}
\def\le{\mathrel{\spose{\lower 3pt\hbox{$-$}}
     \raise 2.0pt\hbox{$\mathchar"13C$}}}

\documentclass[aps,twocolumn]{revtex4}

\usepackage{graphics}

\newcommand {\boom}{{\sc Boomerang }}

\begin{document}

\bibliographystyle{apsrev}

\title{Using the acoustic peak to measure cosmological parameters}
\author{ Neil J. Cornish}
\affiliation{Department of Physics, Montana State University, Bozeman, MT 59717}

\begin{abstract}
Recent measurements of the cosmic microwave background radiation by the \boom and {\sc Maxima}
experiments indicate that the universe is spatially flat. Here some simple
back-of-the-envelope calculations
are used to explain the result. The main result is a simple formula for the angular scale
of the acoustic peak in terms of the standard cosmological parameters:
$\ell \approx 193[1+\frac{3}{5}(1-\Omega_0)+\frac{1}{5}(1-h)
+\frac{1}{35}{\Omega_\Lambda}]$.
\end{abstract}
\pacs{}

\maketitle


As we enter the era of precision cosmology, it gets increasingly difficult to
understand how cosmological parameters are extracted from observational
data. The cosmic microwave background radiation (CMBR) is a prime example.
Anisotropies in the CMBR are influenced by a large number of cosmological
parameters, including, but not limited to; the Hubble constant; the spatial
curvature; the spatial topology; the vacuum energy density; the baryon density; the
number of light neutrinos; and the amplitude and spectral index of the primordial
density perturbations. Using accurate maps of the CMBR, it should be possible
to fix all of these parameters with great precision\cite{dave,bond,css}.
However, with so many parameters and so many physical effects to keep track of,
it is hard to explain how a particular parameter is
extracted from the data\cite{huweb}.
The aim here is to provide a simple explanation of how the microwave background radiation can
be used to measure the spatial curvature. The test was first proposed by
by Doroshkevich, Zel'dovich and Sunyaev\cite{ya}, and has subsequently been
developed by many authors. The most comprehensive
treatment can be found in the work of Hu and White\cite{hu}.

The curvature measurement is based on simple geometry.
If you know the physical size of an object and how far away it is,
then by measuring its angular size you can infer the curvature of space.
Suppose that the object has size $A$ and is a distance $B$ away. In flat space
the angle subtended by the object is given by
\begin{equation}\label{flat}
\alpha = {\rm arccos}\left(1-\frac{A^2}{2B^2}\right) \, .
\end{equation}
However, if the space is negatively curved with radius of curvature $R_c$,
the angle will be given by
\begin{equation}\label{hyper}
\alpha = {\rm arccos}\left(1-\frac{\cosh(A/R_c) -1}{\sinh^2(B/R_c)}\right) \, .
\end{equation}
Notice that (\ref{hyper}) recovers that flat space result (\ref{flat}) in the
limit $R_c \rightarrow \infty$. The expression for a positively curved space
can be obtained by replacing $R_c$ by $iR_c$ in (\ref{hyper}).

To apply the angular size test we need to know the size of a distant object and
how far away it is. When the test is applied
to the microwave background radiation, the role of the distant ``ruler stick''
is played by the size of the sound horizon at last scatter, and the
distance to the object is the radius of the last scattering surface.
Both of these quantities depend on several cosmological parameters, but the spatial
curvature turns out to be the dominant effect.

We can measure the angular size subtended by the sound horizon by looking for
a special feature in the CMBR angular power spectrum. The angular power spectrum is
obtained by ``fourier'' analyzing the CMBR anisotropy pattern. Sound waves in the
photon-baryon fluid with wavelengths roughly twice the size of the sound
horizon at last scatter will have just reached a maximum density contrast when
matter and radiation decouple. As we shall see, these waves have periods that
are long compared to the time taken for matter and radiation to decouple. Thus,
the compression--rarefaction pattern is snap frozen at decoupling. The enhanced
density contrast that occurs on the scale of the sound horizon leads to an
enhanced temperature anisotropy, as the CMBR photons are redshifted by an amount
proportional to the local density. By measuring the scale at which the peak in the
angular power spectrum occurs, we are able to establish the angular size of the
sound horizon. 
Note: The acoustic peaks are {\em density} peaks, not ``Doppler peaks''.
The fluid is at a turn-around point when maximum density contrast is reached,
so the velocity of the baryons, and hence the Doppler shift, is at a minimum.

Before proceeding to show how the angle subtended by the sound horizon is
related to the spatial curvature, a little more non-euclidean geometry is
in order. Consider a geodesic triangle drawn in hyperbolic space. The
law of cosines reads
\begin{eqnarray}
&&\cosh(C/R_c) = \cosh(A/R_c)\cosh(B/R_c)\nonumber \\
&& \hspace*{1in} -\sinh(A/R_c)\sinh(B/R_c)\cos\gamma \, .
\end{eqnarray}
Here $A,B,C$ are the side lengths and $\alpha,\beta,\gamma$ are the opposite
angles. Now suppose that $C=B$, $B \gg A$ and $R_c \gg A$. Using the law of
cosines we have
\begin{equation}\label{alp}
\alpha = \frac{A}{R_c\sinh(B/R_c)} +\dots \, ,
\end{equation}
and
\begin{equation}
\beta = \frac{\pi}{2} - \frac{A}{2R_c\tanh(B/R_c)} +\dots \, .
\end{equation}
The sum of the angles in the triangle is given by
\begin{equation}
\Sigma = \alpha + 2\beta = \pi - \alpha ( \cosh(B/R_c)-1) + \dots \, .
\end{equation}
Thus, the angle sum is less that $180^o$ if space is negatively curved,
greater than $180^o$ if space is positively curved, and equal to $180^o$
if space is flat. In our cosmological setting, $A$ is the size of the
sound horizon, $B$ is the radius of the last scattering surface and
$\alpha$ is the angular scale corresponding to the first acoustic peak.
Space is flat if the angle sum in our cosmic triangle adds to $180^o$.

Turning now from the spatial geometry to the spacetime geometry, the
unperturbed background geometry is described by the Friedman-Robertson-Walker
line element
\begin{equation}
ds^2 = -dt^2 + a(t)^2\left( d\chi^2 + R_c^2 \sinh^2(\chi/R_c)d\Omega^2\right) \, .
\end{equation}
Here $a(t)$ is the scale factor in units where $a_0=1$ today, and $|R_c|$ is
the spatial curvature radius today. The time-time component of Einstein's field
equations reads (in units where $G=c=1$)
\begin{equation}\label{fried}
\left(\frac{\dot{a}}{a}\right)^2 + \frac{1}{a^2 R_c^2} = \frac{8\pi}{3}\rho \, ,
\end{equation}
where $\rho$ is the energy density and a dot denotes $d/dt$. Solving for
$R_c$ we find
\begin{equation}
R_c = \frac{1}{H_0\sqrt{1-\Omega_0}} \, ,
\end{equation}
where $H_0=(\dot{a}/a)_0$ is the Hubble constant and $\Omega_0=\rho_0/\rho_c$
is the total energy density today in units of the critical density $\rho_c=
3H_0^2/8\pi$. Space is negatively curved if $\Omega_0<1$, positively curved
if $\Omega_0>1$ and flat if $\Omega_0=1$. Using conservation of energy-momentum,
the Friedman equation (\ref{fried}) can be rewritten in the useful form
\begin{equation}
H^2=\left(\frac{\dot{a}}{a}\right)^2=H_0^2\left( \frac{\Omega_r}{a^4}+
\frac{\Omega_m}{a^3}+\frac{(1-\Omega_0)}{a^2}+\frac{\Omega_w}{a^{3(1+w)}} \right) \, .
\end{equation}
Here $\Omega_r$, $\Omega_m$ and $\Omega_w$ denote the contributions to the
total energy density from radiation (photons and light neutrinos), non-relativistic
matter (baryons and cold dark matter), and an unclustered dark matter
component with equation of state $p=w\rho$ where $w\leq -1/3$. The unclustered
dark matter takes the form of a cosmological constant when $w=-1$.

Since angles are conformally invariant, we can use
the conformally related static metric
\begin{equation}
d\tilde{s}^2 = -d\eta^2 + d\chi^2 +  R_c^2 \sinh^2(\chi/R_c)d\Omega^2 \, .
\end{equation}
The static (or optical) metric has the nice property that null geodesics in spacetime
correspond to ordinary geodesics in space. The conformal time $\eta$ is
related to the cosmological time $t$ by $d\eta= dt/a$. Our task now is to calculate the
size of the sound horizon at last scatter, and the radius of the last scattering
surface. To leading order, the sound speed in the photon-baryon fluid is equal to
$c_s=1/\sqrt{3}$, so the sound horizon is roughly $1/\sqrt{3}$ times smaller
than the conformal time interval between last scatter and the big bang (or between
last scatter and reheating if we are considering inflationary models). Thus,
the size of the sound horizon is given by
\begin{equation}\label{sh}
\chi_{sh} \simeq  \frac{1}{\sqrt{3}}(\eta_{sls}-\eta_{rh})\simeq \frac{1}{\sqrt{3}}\int_0^{a_{sls}}
\frac{da}{Ha^2} \, .
\end{equation}
The size of the universe at last scatter, $a_{sls}$, is inversely proportional to the redshift of
last scatter, $z_{sls}\approx 1100$. The radius of the surface of last scatter is
equal to the conformal time interval between last scatter and today:
\begin{equation}\label{sls}
\chi_{sls} = \eta_0-\eta_{sls} \simeq \int_0^{1} \frac{da}{Ha^2} \, .
\end{equation}
Using the same approximations used to derive equation (\ref{alp}), we can relate
the angle subtended by the sound horizon, $\theta_{sh}$, to $\chi_{sh}$ and
$\chi_{sls}$:
\begin{equation}\label{main}
\theta_{sh} \approx \frac{\chi_{sh}}{R_c\sinh(\chi_{sls}/R_c)} \, .
\end{equation}

Let us begin with a simple case. Consider a matter dominated universe with
$\Omega_r=\Omega_w=0$. The integrals (\ref{sh}) and (\ref{sls}) yield
\begin{eqnarray}
&&\chi_{sh} = \frac{2}{H_0\sqrt{\Omega_0}\sqrt{3 z_{sls}}}\, , \nonumber \\ \nonumber \\
&&\chi_{sls} = R_c\, {\rm arcsinh}\left( \frac{2 \sqrt{1-\Omega_0}}{\Omega_0}\right) \, ,
\end{eqnarray}
and the angular size of the sound horizon is given by
\begin{equation}
\theta_{sh}=\sqrt{\frac{\Omega_0}{3z_{sls}}} \approx 1^o\, \sqrt{\Omega_0} \, .
\end{equation}
The angle sum in the cosmic triangle is given by
\begin{equation}
\Sigma \approx 180^o -  2^o \, \frac{1-\Omega_0}{\sqrt{\Omega_0}}\, .
\end{equation}
Note that while both $\chi_{sh}$ and $\chi_{sls}$ depend on $H_0$ and $\Omega_0$,
the angles only depend on $\Omega_0$. Thus, in a matter dominated universe, the
position of the first acoustic peak is an excellent measure of the curvature.
Since the CMBR experiments report their results in terms of angular power
spectra, it is conventional to convert the angular scale into its fourier
equivalent, the multipole number $\ell \simeq \pi/\theta$. For a matter
dominated universe, the first acoustic peak in the angular power spectrum
is located at $\ell_{peak}\approx \ell_{sh}\simeq  180/\sqrt{\Omega_0}$.

For more realistic cosmological models, with both radiation and
multi-component dark matter, the integrals (\ref{sh}) and (\ref{sls})
can not be evaluated in terms of simple functions. Approximate forms
can be found as a power series expansions in the quantities $\Omega_w/\Omega_m$,
$\Omega_c/\Omega_m$ and $z_{sls}/z_{eq}$, where $\Omega_c = 1-\Omega_0$ is
a measure of the curvature and $z_{eq}\simeq \Omega_m/\Omega_r$ denotes the
redshift of matter-radiation equality. To leading order we have
\begin{eqnarray}\label{fo}
&&\chi_{sh} \simeq \frac{2}{H_0\sqrt{\Omega_m}
\sqrt{3 z_{sls}}}\left(\sqrt{1+\frac{z_{sls}}{z_{eq}}}-
\sqrt{\frac{z_{sls}}{z_{eq}}}\right)
\, , \nonumber \\ 
\nonumber \\
&&  {\rm and} \\ \nonumber \\
&&\chi_{sls} \simeq \frac{2}{H_0\sqrt{\Omega_m}}\left(1-\frac{1}{6}\frac{\Omega_c}{\Omega_m}
-\frac{1}{2(1-6w)}\frac{\Omega_w}{\Omega_m} + \dots \right)\,  . \nonumber
\end{eqnarray}
The expression for the radius of the last scattering
surface, $\chi_{sls}$, is good to within $10\%$
for $|\Omega_c/\Omega_m| \lta 2$ and $|\Omega_w/\Omega_m| \lta 2$
- see the appendix for details.
The quantity $z_{sls}/z_{eq}$ that appears in (\ref{fo}) is well approximated by
\begin{equation}
\frac{z_{sls}}{z_{eq}}\simeq \frac{1}{24\, \Omega_m h^2} \, ,
\end{equation}
where $h$ is the Hubble constant in units of 100 km s$^{-1}$ Mpc$^{-1}$.
In order to find a simple expansion for $\chi_{sh}$, we can choose either
$h$ or $\Omega_m h^2$ as a free parameter and expand the quantity
$(\sqrt{1+z_{sls}/z_{eq}}-\sqrt{z_{sls}/z_{eq}})$ as
\[
\frac{1}{\sqrt{3}}\left(1+
\frac{(8\Omega_m h^2-1)}{4}-\frac{9(8\Omega_m h^2-1)^2}{64} + \dots\right)\, ,
\]
or
\[
\sqrt{\frac{2}{3}} \left(1-\frac{(1-\Omega_m)}{10}
-\frac{(1-h)}{5} + \dots\right)\, .
\]
The first version works best when $\Omega_m$ and $h$ are close to
the currently favored values of $\Omega_m=0.3$ and $h=0.65$. The
second version works best if $\Omega_m\simeq 1$ and $h\simeq 1$.
Putting everything together in (\ref{main}) we find
\begin{eqnarray}\label{shap}
&&\theta_{sh} \simeq \frac{1}{3\sqrt{z_{sls}}}\left(
1-\frac{1}{2}\frac{\Omega_c}{\Omega_m}
+\frac{1}{2(1-6w)}\frac{\Omega_w}{\Omega_m} \right. \nonumber \\
&& \hspace*{1.3in} \left.
 + \frac{(8\Omega_m h^2-1)}{4}+
\dots \right) \, ,
\end{eqnarray}
or, specializing to the case $w=-1$ and $\Omega_m\simeq 1$, we find
\begin{equation}\label{simple}
\theta_{sh} \simeq \frac{\sqrt{2}}{3\sqrt{z_{sls}}}\left(
1-\frac{3\Omega_c}{5}
 -\frac{\Omega_\Lambda}{35}
 -\frac{(1-h)}{5}+
\dots \right) \, .
\end{equation}
The above expressions for $\theta_{sh}$ give a good qualitative picture of
how the various cosmological parameters affect the location of the first acoustic
peak. We see that the peak position is mainly determined by the curvature,
and only weakly dependent on the value of the Hubble constant. The peak position
is largely insensitive to the value of the cosmological constant.

The angle subtended by the sound horizon is
smaller in a negatively curved universe and larger in a positively curved
universe. The angles in the triangle formed by the sound horizon and the Earth
sum to
\begin{equation}
\Sigma \simeq 180^{o}-1.15^{o}\frac{\Omega_c}{\Omega_m}\left(1
-\frac{\Omega_c}{2\Omega_m}-\frac{\Omega_\Lambda}{14\Omega_m}+\frac{8\Omega_m h^2-1}{4}
\right).
\end{equation}
Neglecting photon self-gravity, sound waves in the photon-baryon fluid obey a
simple harmonic oscillator equation, and the position of the first acoustic
peak corresponds to the angular size
of the sound horizon $\ell_{sh} \simeq \pi/\theta_{sh}$. However, when
photon self-gravity is included, the oscillator equation gains an anharmonic term
that shifts the position of the first few peaks. Taking this
into account, and using standard isentropic initial conditions, and neglecting
Silk damping, the temperature fluctuations vary as a function
of scale as~\cite{hu}
\begin{equation} {\cal T} \approx {\rm constant} -\cos(\ell \theta_{sh})+
\frac{1}{\ell \theta_{sh}}\sin(\ell \theta_{sh}) \, .
\end{equation}
Solving for the position of the first peak, we find
\begin{equation}
\ell_{peak}\approx 0.873\, \frac{\pi}{\theta_{sh}} \, ,
\end{equation}
so that for $\Omega_m\approx 1$
\begin{equation}\label{fnl}
\ell_{peak}\approx 193\left(1+\frac{3\Omega_c}{5}+\frac{\Omega_\Lambda}{35}+\frac{(1-h)}{5}+
\dots \right) \, .
\end{equation}
The recent {\sc Mat }$\!$\cite{amber}, \boom$\!$\cite{boom} and {\sc Maxima}$\!$\cite{max}
results locate the first acoustic peak at $\ell \approx 200$, $\ell = 197 \pm 6$
and $\ell \approx 220$, respectively. Using the \boom results, and allowing $h$ and
$\Omega_\Lambda$ to vary freely over the range $0.5 \leq h \leq 0.8$ and
$0\leq \Omega_{\Lambda}<0.8$, our approximate formula
(\ref{fnl}) yields a best fit value of $\Omega_0 = 1.07 \pm 0.1$. This result is
consistent with the universe being spatially flat, and agrees with the
detailed analysis\cite{bond2} of the \boom data. Since the curvature
is the dominant effect in fixing the location of the acoustic peak, we are
able to get a good fix on $\Omega_0$ despite having only one equation for
three unknowns $(H_0,\Omega_0,\Omega_\Lambda)$.
In conclusion, simple analytic formulas can be found that
give good qualitative, and decent quantitative, insight into how
the CMBR observations are used to fix the spatial curvature.

\section*{Acknowledgements}
I would like to thank David Spergel and Wayne Hu for patiently and expertly
answering all my questions.

\section*{appendix}
We made two major approximations in arriving at equation (\ref{shap}). The first
was to treat the sound speed as a constant, when a more
accurate approximation would be to set $c_s=1/\sqrt{3(1+\xi)}$,
where $\xi = 3a\, \Omega_b/4\Omega_\gamma$ is the baryon-photon momentum
density ratio\cite{hu}. Keeping the next to leading term in $\xi_{sls}$, the
size of the sound horizon is given by
\begin{eqnarray}\label{shnext}
&&\chi_{sh} = \frac{2}{H_0\sqrt{\Omega_m}\sqrt{3 z_{sls}}}\left[
\, \left(\sqrt{1+\frac{z_{sls}}{z_{eq}}}-
\sqrt{\frac{z_{sls}}{z_{eq}}}\right)\right. \nonumber \\
\nonumber \\
&& \hspace*{0.2in} \left. -\frac{\xi_{sls}}{6}\left(\sqrt{1+\frac{z_{sls}}{z_{eq}}}-
2\frac{z_{sls}}{z_{eq}}\left(\sqrt{1+\frac{z_{sls}}{z_{eq}}}-
\sqrt{\frac{z_{sls}}{z_{eq}}}\right)\right)\right] \nonumber \\
\end{eqnarray}
Assuming that there are three light neutrino species,
the quantity $\xi_{sls}$ can be written as
\begin{equation}
\xi_{sls}=\frac{3}{4}\left(1+\frac{21}{8}\left(\frac{4}{11}\right)^{4/3}\right)
\left(\frac{\Omega_b}{\Omega_m}\right)\left(\frac{z_{eq}}{z_{sls}}\right) \, .
\end{equation}
For reasonable values of the cosmological parameters, we find $\xi_{sls} \lta 1$, so we
can negelect the second term in equation (\ref{shnext}).

\begin{picture}(0,0)
\put(-5,-100){$\Omega_c$}
\put(120,-220){$\Omega_\Lambda$}
\end{picture}
\begin{figure}[h]
\vspace{2.9in}
\includegraphics{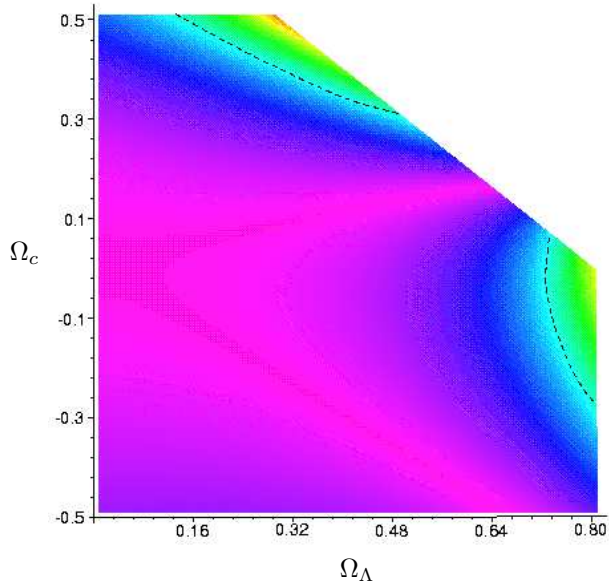}
\caption{\protect\small The fractional error in the first order approximation
to angular size distance to the surface of last scatter. The dashed lines mark
contours of $8\%$ error. The error is considerably less than $8\%$ across most of
parameter space. The missing corner corresponds to the portion of parameter space
where $\Omega_m=(\Omega_0-\Omega_\Lambda)<0.2$.} 
\end{figure}

The second major approximation was to expand the expression for
$r_{sls}=R_c\sinh(\chi_{sls}/R_c)$ in terms of
$\Omega_c/\Omega_m$ and $\Omega_w/\Omega_m$:
\begin{eqnarray}
&& r_{sls} = \frac{2}{H_0\sqrt{\Omega_m}}\left(
1 + \frac{1}{2}\frac{\Omega_c}{\Omega_m}-\frac{1}{2(1-6w)}\frac{\Omega_w}{\Omega_m}\right.
\nonumber \\ \nonumber \\
&& \hspace*{0.4in}  -\frac{1}{8}\left(\frac{\Omega_c}{\Omega_m}\right)^2+
\frac{3}{8(1-12w)}\left(\frac{\Omega_w}{\Omega_m}\right)^2
\nonumber \\ \nonumber \\
&& \hspace*{0.2in} \left. -
\frac{3-2w}{4(1-2w)(1-6w)}\left(\frac{\Omega_c}{\Omega_m}\frac{\Omega_w}{\Omega_m}\right)
+\dots \right) . \\
\nonumber 
\end{eqnarray}
So long as $\Omega_m > \Omega_c$ and $\Omega_m > \Omega_w$, the higher order terms
can safely be neglected. Figure 1. shows the percentage error in the first order
truncation of $r_{sls}$ as compared to a full numerical evaluation. The fractional error is
less than $8\%$ across a wide portion of parameter space, including the interesting
region around $(\Omega_c,\Omega_\Lambda)=(0,0.7)$.

Our final task is to show that the period of the wave responsible for the
first acoustic peak is large compared to the time taken for matter and
radiation to decouple. If this were not the case, the anisotropy would
not be frozen in and the acoustic peak would be washed out. The conformal period
of the wave is given by $T\simeq 2\pi \chi_{sh}/c_s$ and the conformal
time interval taken to decouple is $\Delta \eta = \eta(z_{sls})-\eta(z_{sls}+\Delta z)$,
where $\Delta z\approx 300$ is the redshift interval for decoupling.
The ratio of $T$ to $\Delta \eta$ is given by
\begin{equation}
\frac{T}{\Delta\eta} \simeq 4\pi \left(\frac{z_{sls}}{\Delta z}\right)\left(
1+\frac{z_{sls}}{z_{eq}}-\sqrt{\frac{z_{sls}}{z_{eq}}\left(1+\frac{z_{sls}}{z_{eq}}
\right)} \right) \, .
\end{equation}
For reasonable cosmological parameters, we find $T/\Delta \eta \gta 30$, which tells us
that the acoustic waves are effectively snap frozen when matter and radiation decouple.

\end{document}